\documentclass[12pt]{iopart}
\usepackage{graphicx}
\usepackage{bm}

\usepackage{here}
\usepackage{iopams}  
\begin{document}

\title[]{Raman scattering for triangular lattices spin-$\frac{1}{2}$ Heisenberg antiferromagnets}

\author{F. Vernay$^{1,2}$, T. P. Devereaux$^{1,2}$ and M. J. P. Gingras$^{1,3}$}

\address{$^1$Department of Physics, University of Waterloo, Waterloo, ON, N2L3G1, Canada\\
$^{2}$PITP, University of British Columbia, Vancouver, BC, V6T1Z1, Canada\\
$^3$Department of Physics and Astronomy, University of Canterbury, Christchurch 8020, New-Zealand}
\ead{vernay@lorax.uwaterloo.ca}
\begin{abstract}
Motivated by various spin-1/2 compounds like Cs$_2$CuCl$_4$ or $\kappa$-(BEDT-TTF)$_2$Cu$_2$(CN)$_3$,
we derive a Raman-scattering operator {\it \`a la} Shastry and Shraiman for various geometries.
For T=0, the exact spectra is computed by Lanczos algorithm for finite-size clusters. 
We perform a systematic investigation as a function of $J_2/J_1$, the exchange constant ratio:
ranging from $J_2=0$, the well known square-lattice case, to $J_2/J_1=1$ the isotropic triangular lattice.
We discuss the polarization dependence of the spectra and show how it can be used 
to detect precursors of the instabilities of the ground state against quantum fluctuations.
\end{abstract}

\section{Introduction} 
Highly frustrated magnetic systems are highly susceptible to quantum
spin fluctuations and instabilities towards competing ground
states. The triangular Heisenberg antiferromagnet
with spin $S=1/2$ constitutes a paradigm of that class of systems.
In that context, it is interesting that experimental investigations 
 of the Cs$_2$CuCl$_4$ \cite{coldea} and
$\kappa$-(BEDT-TTF)$_2$Cu$_2$(CN)$_3$ \cite{shimizu} materials,
which can be both described as
a first approximation by a 
triangular lattice with
spatially anisotropic exchange couplings, indicate exotic behaviors.
Exotic behaviors include either the realization of a spin-liquid 
ground state or a magnetically ordered phase with a magnetic excitation 
dispersion strongly renormalized compared to the classical 
spin-wave. Indeed, recent numerical studies based on series expansion \cite{zheng}
have found that frustration manifests itself rather directly
in the spin excitations of archetype models of two-dimensional
frustrated spin systems.
A softening of the magnon excitations in a broad region of
the reciprocal ${\bm q}$ space is observed as a quantity, 
$f$, which parametrize the level of frustration, is increased.
Since, within a semi-classical picture,
the effect of frustration is to bring about the effect
of competing ground states, it is expected that
the spin excitations out of a semi-classical long-range ordered
ground state are direct tell-tale indicators of the presence
of frustrating interactions. In this paper we explore the possibility
that frustration can also be investigated via the two-magnon 
density of states at zero wavevector measured by polarized inelastic 
magnetic Raman scattering.\\
The rest of this paper is organized as follows: in the first  
Section we present the derivation of the scattering operator, then we 
discuss its form for various polarizations. In the last part we present 
the associated Raman spectra obtained by exact-diagonalization of finite-size 
clusters. 
\section{Scattering operator} 
\begin{figure}[H]
\begin{center}
\includegraphics*[width=9cm,angle=0]{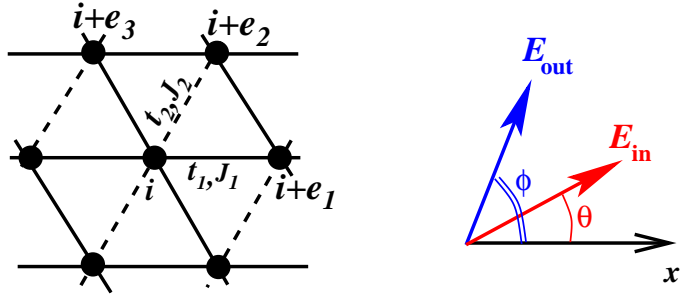}
\caption[figure3]{\label{pola}
Geometry of the lattice and polarization vectors. 
{\it Left:} The dashed lines represent the coupling $J_2$, 
the thick lines are for the exchange coupling $J_1$. 
{\it Right:} Definition of the polarization vectors by two angles $\phi$ and $\theta$.}
\end{center}
\end{figure}

We first consider the Hubbard model on the anisotropic triangular lattice:
\begin{equation}\label{hamilt}
{\mathcal H}={\mathcal H}_{K}+{\mathcal H}_{U}=\sum_{\langle i,j\rangle ,\sigma} t_{ij} 
\left(c^\dagger_{i,\sigma} c_{j,\sigma} + h.c.\right)+ U\sum_{i} n_{i,\uparrow} n_{i,\downarrow}
\end{equation}
where $t_{ij}$ is the hopping from a site $i$ to a neighboring site $j$, $\sigma$ is the 
spin degree of freedom, and $c$ and $c^\dagger$ are the usual 
annihilation and creation operators. 
For the case considered here, there
are two distinct hopping integrals,
$t_1$ and $t_2$, along the corresponding directions shown in Fig.~\ref{pola}.\\ 
At half-filling and in the large-$U$ limit, the system 
is described (to second order in $t/U$) by an effective spatially anisotropic 
Heisenberg Hamiltonian:
\begin{equation}\label{hamiltheis}
{\mathcal H}_{\rm eff}=\sum_{\langle i,j \rangle} J_{ij} {\bf S}_i\cdot{\bf S}_j
\end{equation}
where the different exchange coupling along each direction is due to the difference
in the hopping amplitude, {\it i.e} $J_1=4t^2_1/U$~, $J_2=4t^2_2/U$.\\ 
Raman scattering consists of an incoming photon (of energy $\omega_i$) scattered into an 
outgoing photon (of energy $\omega_f$), involving different manifold of electronic states having zero or 
one double-occupancy. 
These transitions depend on the polarizations (referred as 
{\bf E}$_{\rm in}$ and ${\bf E}_{\rm out}$) 
of the incoming and outgoing photons. 
Following the early work of Fleury and Loudon \cite{loudon} and Shastry and Shraiman \cite{shastry}, we 
derive an effective scattering spin Hamiltonian describing this problem.  
The $\alpha$-component of the electronic hopping current operator is:
\begin{equation}\label{curr}
j^\alpha({\bf q})=i\sum_{{\bf r},{\bm \nu}=\pm({\bf e_1,e_2,e_3})}
\left(\frac{\partial \epsilon_{\bf k}}{\partial k^\alpha}\right) 
e^{i {\bf q}\cdot({\bf r}+\frac{{\bm\nu}}{2})}
\left[c^\dagger_\sigma({\bf r}+{\bm\nu})c_\sigma({\bf r})-
c^\dagger_\sigma({\bf r})c_\sigma({\bf r}+{\bm\nu})\right]
\end{equation}
where ${\bf q}={\bf k}_f-{\bf k}_i$ is the momentum transfer, and $\epsilon_{\bf k}$ 
the band-energy. In what follows we take the most general set of polarization vectors, as  
shown in the right panel of Fig.~\ref{pola}:
\begin{equation}
\left\{\begin{array}{rcl}
\cos\phi\ \hat{x}&+&\sin\phi\ \hat{y}\\
\cos\theta\ \hat{x}&+&\sin\theta\ \hat{y}\\
\end{array}\right.
\end{equation}
The Raman scattering process involves an energy transfer $\Omega=\omega_f-\omega_i$, 
the momentum transfer being set to zero. Therefore, we restrict ourselves to the assumption 
${\bf k}_f\approx{\bf k}_i\approx 0$, and we have:
\begin{equation}
\begin{array}{c}
{\bf j}\cdot{\bf E}_{\rm in} = i\cos\phi
\left[-t_1\left(c^\dagger_r c_{r+{\bf e}_1}- h.c\right)\right.
\left.-\frac{t_2}{2}\left(c^\dagger_r c_{r+{\bf e}_2}- h.c\right)\right.\\\left.+
\frac{t_1}{2}\left(c^\dagger_r c_{r+{\bf e}_3}- h.c\right)\right]\\
-i\frac{\sqrt{3}}{2}\sin\phi\left[t_2\left(c^\dagger_r c_{r+{\bf e}_2}- h.c\right)+
t_1\left(c^\dagger_r c_{r+{\bf e}_3}- h.c\right)
\right]\\
\ \\ 
{\bf j}\cdot{\bf E}_{\rm out} = i\cos\theta 
\left[-t_1\left(c^\dagger_r c_{r+{\bf e}_1}- h.c\right)\right.
\left.-\frac{t_2}{2}\left(c^\dagger_r c_{r+{\bf e}_2}- h.c\right)\right.\\\left.+
\frac{t_1}{2}\left(c^\dagger_r c_{r+{\bf e}_3}- h.c\right)\right]\\
-i\frac{\sqrt{3}}{2}\sin\theta\left[t_2\left(c^\dagger_r c_{r+{\bf e}_2}- h.c\right)+
t_1\left(c^\dagger_r c_{r+{\bf e}_3}- h.c\right)
\right]\\
\end{array}
\end{equation}
The Raman scattering operator is given by the second order formula, 
$|i\rangle$ and $|f\rangle$ being respectively the initial and final 
state, $|\mu\rangle$ being an intermediate state:
\begin{equation}\label{scat}
\begin{array}{rcl}
\langle f|M_r|i\rangle &=& \sum_\mu \left[
\frac{\langle f|{\bf j}\cdot{\bf E}_{\rm out}|\mu\rangle
\langle\mu|{\bf j}\cdot{\bf E}_{\rm in}|i\rangle}
{\epsilon_\mu-\epsilon_i-\omega_i} \right.
+\left.
\frac{\langle f|{\bf j}\cdot{\bf E}_{\rm in}|\mu\rangle
\langle\mu|{\bf j}\cdot{\bf E}_{\rm out}|i\rangle}
{\epsilon_\mu-\epsilon_i+\omega_f}\right]
\end{array}
\end{equation}
Following the same algebra steps as in \cite{shastry}, and restricting 
$|i\rangle$ and $|f\rangle$ to the manifold of singly occupied states, and 
intermediate states $|\mu\rangle$ to the manifold of one double occupancy, 
we use the identity, 
$\frac{1}{4}-{\bf S}_i\cdot{\bf S}_j=\sum_{\sigma,\sigma'}
\frac{1}{2}c^\dagger_{i,\sigma} c_{j,\sigma} c^\dagger_{j,\sigma'} c_{i,\sigma'}$, 
and obtain the scattering operator in terms of spin operators. We note that within 
our approach the total scattering operator only contains terms of the form:
\begin{equation}\label{scat2}
{\mathcal O}_\nu\propto \frac{t_\nu^2}{U} {\bf S}_i\cdot{\bf S}_{i+{\bf e}_\nu}
\end{equation}
From Eq. (\ref{scat}) we find that the scattering Hamiltonian 
prefactors which will be in front of Eq. (\ref{scat2}) 
depend on the polarization vectors orientation. 
These prefactors are proportional to 
the exchange coupling $J$, therefore, in its general form, 
the total scattering operator will depend on both $J_1$ and $J_2$.
\section{Polarizations}    
The two angles $\phi$ and $\theta$ 
with respect to the $x$-axis (see Fig. \ref{pola})
define the polarizations involved in the scattering process.
Therefore the scattering operator depends on a projector ${\mathcal P}_\nu(\theta,\phi)$ that defines the 
polarization set-up. The Raman operator takes the form:
\begin{equation}
\begin{array}{c}
{\mathcal H}_{LF}(\theta,\phi) \propto \sum_i \left\{J_1 {\bf S}_i\cdot{\bf S}_{i+{\bf e}_1} \cos\theta\cos\phi\right.\\
 \ \\
+ J_2 {\bf S}_i\cdot{\bf S}_{i+{\bf e}_2}\left[\cos(\theta+\phi)+\sqrt{3}\sin(\theta+\phi)+4\sin\theta\sin\phi\right]\\
 \ \\
+ \left.J_1 {\bf S}_i\cdot{\bf S}_{i+{\bf e}_3}\left[\cos(\theta+\phi)-\sqrt{3}\sin(\theta+\phi)+4\sin\theta\sin\phi\right]
\right\}
\end{array}
\end{equation}
Which can be written in the compact form:
\begin{equation}\label{operator}
{\mathcal H}_{LF}(\theta,\phi)
\propto \sum_{i,\nu} J_\nu {\mathcal P}_\nu(\theta,\phi)
{\bf S}_i\cdot{\bf S}_{i+{\bf e}_\nu}
\end{equation}
In order to compare with the square-lattice case, we now focus 
on the following polarization geometries:
\begin{equation}\label{a1gb1g}
\begin{array}{rcl}
{\mathcal H}_{LF}(\frac{5\pi}{6},-\frac{\pi}{6})
&\propto& \sum_{i} J_1 \left[{\bf S}_i\cdot{\bf S}_{i+{\bf e}_1}+{\bf S}_i\cdot{\bf S}_{i+{\bf e}_3}\right]
\\
{\mathcal H}_{LF}(\frac{5\pi}{6},\frac{\pi}{3})
&\propto& \sum_{i} J_1 \left[{\bf S}_i\cdot{\bf S}_{i+{\bf e}_1}-{\bf S}_i\cdot{\bf S}_{i+{\bf e}_3}\right]
\end{array}
\end{equation}
When the diagonal bond $J_2$ is $0$, the first line of Eq. (\ref{a1gb1g}) is the A$_{1g}$ Raman operator, 
while the second line gives the B$_{1g}$ Raman operator, both on a square lattice. 
Most importantly, we note that the scattering operators depend on the ratio $J_1/J_2$
if one takes:
\begin{equation}\label{a1gb1g-J2}
\begin{array}{rcl}
{\mathcal H}_{LF}(\frac{\pi}{6},-\frac{5\pi}{6})
&\propto& \sum_{i} \left[J_1 {\bf S}_i\cdot{\bf S}_{i+{\bf e}_1}+J_2{\bf S}_i\cdot{\bf S}_{i+{\bf e}_2}\right]
\\
{\mathcal H}_{LF}(\frac{\pi}{6},\frac{2\pi}{3})
&\propto& \sum_{i}\left[J_1{\bf S}_i\cdot{\bf S}_{i+{\bf e}_1}-J_2{\bf S}_i\cdot{\bf S}_{i+{\bf e}_2}\right]
\end{array}
\end{equation}
We would like to attract the attention of the reader to the fact that the 
standard parallel polarization does not give a 
straightforward A$_{1g}$-like scattering operator as it is the 
case for the square lattice, but a linear combination 
of Eqs. \ref{a1gb1g} and \ref{a1gb1g-J2}. 
\section{Results}          
\begin{figure}[ht]
\begin{center}
\includegraphics*[width=14.5cm,angle=0]{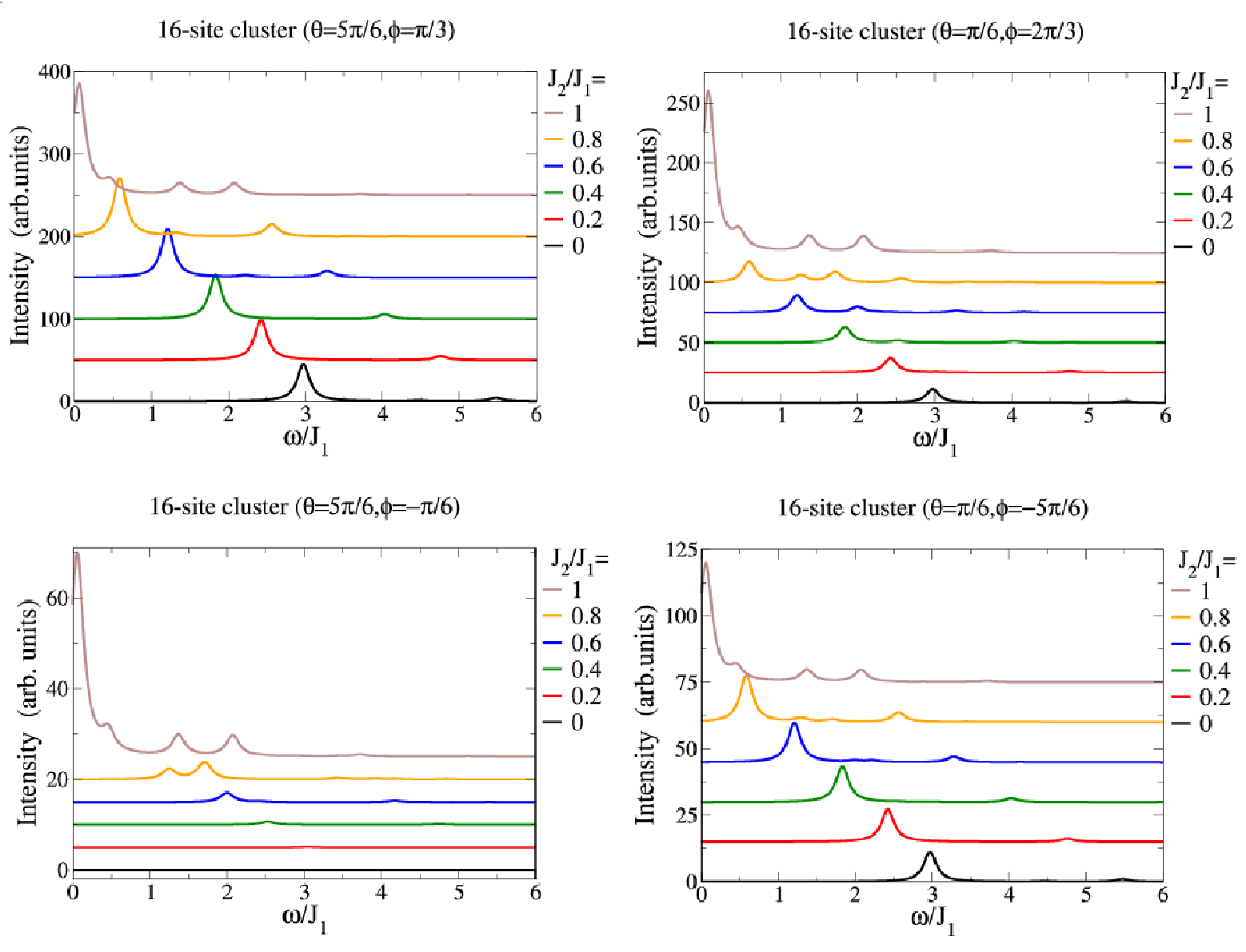}
\caption[figure3]{\label{raman}
Exact diagonalizations on a 16-site cluster.
Raman spectra for various values of $(\theta,\phi)$ and $J_2/J_1$.
The bottom (top) left panel would correspond to $A_{1g}$
($B_{1g}$) polarization on the
square lattice in the case $J_2=0)$.}
\end{center}
\end{figure}

As a first step in exploring the Raman scattering in the
Heisenberg $S=1/2$
anisotropic triangular
lattice
antiferromagnet,
 we  
perform exact-diagonalizations for the Raman operator Eq. (\ref{operator}). 
While Raman scattering studies on
 triangular lattice compounds like cobaltites have been carried out \cite{lemmens},
these have focused on phonons, and  
there are yet no systematic study of polarization dependent electronic Raman spectra.
Hence, our aim at this stage is not to obtain a quantitative description of the scattering spectra for 
a specific material, but rather 
identify the key qualitative features emerging in the Raman spectrum of the anistropic triangular lattice, 
possibly to motivate
Raman studies and to connect with recent neutron data \cite{coldea}, NMR \cite{shimizu} and 
angle-resolved photoemission \cite{hasan}. 
Motivated by the successes of exact-diagonalization to identify the essential qualitative
features of the Raman spectrum on the square lattice \cite{sandvik,freitas}, and 
in order to compare with these well-known results, we explore in this study  
the behaviour of a 16-site cluster as the frustrating $J_2$ coupling is increased from zero. 
The results are summarized in Fig. \ref{raman}. 
The general trend observed in these plots is that a softening
progressively develops as the system becomes more frustrated. 
The bottom left panel of Fig. \ref{raman} shows, for
$J_2=0$, the well known zero A$_{1g}$ Raman scattering 
on a square lattice, since the scattering operator
commutes with the Hamiltonian \cite{freitas}. 
As $J_2$ increases non-commuting contribution 
to the scattering intensity increases and the spectrum 
develops more structure. For the bottom right panel, 
a weak intensity if observe because the scattering operator 
does not commute with the Hamiltonian even for $J_2=0$.\\
From the Hamiltonian in Eq.~\ref{hamiltheis}, 
it is obvious that the ground state will depend on the ratio $J_2/J_1$, 
ranging from a N\'eel-like order on the square lattice when $J_2=0$, 
to a three sublattice long-range order for $J_2=J_1$ \cite{weihong,chung,bernu}. 
These different type of order are characterized by different 
magnon dispersions, therefore the Raman 
spectra is expected to depend on $J_2/J_1$. 
In a recent paper, Zheng {\it et al.} \cite{zheng} showed, using series expansions, 
that the magnon dispersion for the anisotropic triangular lattice Heisenberg model 
exhibits a {\it roton-like} minimum. This local minimum sits at the $(\pi,0)$ point for a 
square lattice with one diagonal bond and is getting softer as the $J_2$-diagonal 
frustrating coupling increases. In the case of the standard square lattice, 
for the B$_{1g}$ channel with crossed polarizations, the electromagnetic field couples 
to excitations along the $(\pi,0)$ direction. We note from \cite{zheng} that this softening 
is a multi-magnon process that cannot be captured to lowest order in a 1/S expansion.
 However, since 
we are using exact-diagonalization which, by its nature, treats processes for the length
scale considered exactly, we  are able to detect indications of this softening. 
For specificity, consider the B$_{1g}$ channel in the top left-hand panel of Fig. 2. 
A signature of the softening of the magnon excitation is clearly manifest as the 
frustration ratio $f=J_2/J_1$ is increased from zero, with a shift of the spectral weight,  
and the main peak moving from $\omega \approx 3J_1$ when $J_2=0$ to $\omega \approx 0.6J_1$ 
when $J_2=0.8J_1$. 
\section{Conclusions and perspectives}  
The results presented here show how frustration can dramatically 
alter the Raman spectrum of an otherwise non-frustrated system. 
In particular, we observe a 
frustration-driven {\it spectral downshift}, the landmark feature of frustrated
systems. This shift is another indicator of 
the frustration-induced softening of the magnon dispersion recently predicted by others. 
From a strictly theoretical point of view, the analysis above sets the stage for a more 
complete approach. Some exact-diagonalizations on larger cluster and a spin-wave approach 
including magnon-magnon interactions, along the lines pursued for the square 
lattice \cite{CG,CF}, are currently being carried out.
\section*{Acknowledgments}
Support for this work was provided by NSERC
of Canada and the Canada Research Chair Program (Tier I) (M.G.),  the Canada Foundation for Innovation,
the Ontario Innovation Trust, and the Canadian Institute for Advanced Research (M.G.).
M.G. acknowledges the University of Canterbury for an Erskine Fellowship
and thanks the Department of Physics and Astronomy at the University of Canterbury
for their hospitality where part of this work was completed.

\end{document}